\documentclass[12pt,a4paper]{article}
\usepackage[utf8]{inputenc}
\usepackage{amsmath}
\usepackage{amsfonts}
\usepackage{amssymb}
\usepackage{graphicx, datetime, abstract}
\usepackage{float}
\usepackage{fullpage}
\usepackage{caption}
\usepackage{subcaption}
\usepackage{todonotes}
\usepackage{setspace}

\begin{document}
	\title{
			\begin{flushright}\ \vskip -2cm {\small{\em DCPT-15/13}}\end{flushright}
		Phases and approximations of baryonic popcorn in a\\low-dimensional analogue of holographic QCD}
	\author{Matthew Elliot-Ripley$^*$\\[10pt]
		{\em \normalsize Department of Mathematical Sciences, }\\{\em \normalsize Durham University, Durham, DH1 3LE, U.K.}\\[10pt]
		{\normalsize $^*$ m.k.i.d.elliot-ripley@durham.ac.uk}}
		\date{July 2015}
	\maketitle
	\vspace{15pt}
	\begin{abstract}
The Sakai-Sugimoto model is the most pre-eminent model of holographic QCD, in which baryons correspond to topological solitons in a five-dimensional bulk spacetime. Recently it has been shown that a single soliton in this model can be well approximated by a flat-space self-dual Yang-Mills instanton with a small size, although studies of multi-solitons and solitons at finite density are currently beyond numerical computations. A lower-dimensional analogue of the model has also been studied in which the Sakai-Sugimoto soliton is replaced by a baby Skyrmion in three spacetime dimensions with a warped metric. The lower dimensionality of this model means that full numerical field calculations are possible, and static multi-solitons and solitons at finite density were both investigated, in particular the baryonic popcorn phase transitions at high densities. Here we present and investigate an alternative lower-dimensional analogue of the Sakai-Sugimoto model in which the Sakai-Sugimoto soliton is replaced by an $O(3)$-sigma model instanton in a warped three-dimensional spacetime stabilised by a massive vector meson. A more detailed range of baryonic popcorn phase transitions are found, and the low-dimensional model is used as a testing ground to check the validity of common approximations made in the full five-dimensional model, namely approximating fields using their flat-space equations of motion, and performing a leading order expansion in the metric.
	\end{abstract}
	
	\newpage 

\clearpage
\graphicspath{{./WigglyPopcornPics/}}
\section{Introduction}
The most pre-eminent model of holographic QCD is the Sakai-Sugimoto model \cite{Sakai:2004cn}, \cite{Sakai:2005yt}. The model is a top-down string embedding, which can be reduced to a five-dimensional Yang-Mills-Chern-Simons theory in the holographic limit (in which the number of quark colours is large). As in other holographic QCD models, topological solitons in the bulk, namely Yang-Mills instantons modified by a Chern-Simons term, are dual to (extended) Skyrmions on the boundary. The topological charge shared by these configurations is identified with baryon number.

At large t'Hooft coupling the size of the bulk instantons becomes small compared to the curvature scale of the spacetime, and the charge-1 instanton of the Sakai-Sugimoto model can be well approximated by a flat-space self-dual Yang-Mills instanton \cite{Hong:2007kx}, \cite{Hata:2007mb}. This fact was recently confirmed by numerical computations \cite{Bolognesi:2013nja} by exploiting extra symmetries of the charge-1 instanton, thus reducing the dimensionality of the problem to a 2-dimensional one. However, solitons in the Sakai-Sugimoto model of higher charges do not possess such exploitable symmetries, and such solutions are not well understood, neither analytically nor numerically.

Also beyond the realm of current numerical study are finite density soliton configurations of the Sakai-Sugimoto model. Finite density solitons correspond to cold, dense QCD and it is expected that, in the holographic limit, cold nuclear matter should become a crystalline solid. Two competing theories of what happens to Sakai-Sugimoto solitons at finite density have been proposed, namely the dyon salt \cite{Rho:2009ym} and baryonic popcorn \cite{Kaplunovsky:2012gb}, \cite{Kaplunovsky:2015zsa}. The main focus of this paper will be the popcorn model, in which a lattice of solitons with a finite number of layers in the holographic direction undergoes a series of first-order phase transitions, called \emph{popcorn transitions}, with increasing density. Each of these transitions results in a new lattice with additional layers in the holographic direction. However, studying the dyon salt and popcorn phases is complicated both analytically and numerically. 

To try to gain some further insight into the behaviour of the Sakai-Sugimoto solitons, a lower-dimensional analogue of the model was recently studied \cite{Bolognesi:2013jba}. Since instantons of $O(3)$-sigma models in two space dimensions are well known natural analogues of Yang-Mills instantons in four space dimensions an $O(3)$-sigma model was studied, coupled to a baby Skyrme term in order to stabilise the solitons against the shrinking due to the curvature of space (a role played by the Chern-Simons term in the full Sakai-Sugimoto model). The advantage of the low-dimensional theory is that numerical computations become more viable, and both isolated soliton solutions and finite-density configurations were found. It was found that the dyon salt phase was not energetically favourable, and that beyond a critical density the system undergoes a first-order popcorn transition (albeit one qualitatively different to that predicted in \cite{Kaplunovsky:2012gb}, \cite{Kaplunovsky:2015zsa}) in which a single chain of solitons splits suddenly into two aligned chains separated in the holographic direction. At even higher densities another popcorn transition was found where the double chain splits into a triple chain, and it was conjectured that this procedure would continue so that at higher densities the configuration begins to resemble a portion of a two-dimensional crystal. It was also shown that these chain solutions could be well approximated by periodic chains of flat-space instantons.

However, an alternative low-dimensional analogue of the Sakai-Sugimoto model can be obtained by replacing the baby Skyrme term with a vector meson term in which a new vector field couples to the topological current of the $O(3)$-sigma fields. This would be a closer analogue to the full Sakai-Sugimoto model with the vector field playing the role of the $U(1)$ gauge field that couples to the topological current through the Chern-Simons term. Such a model was studied in flat space in \cite{Foster:2009rw}, and found to be closely related to the baby Skyrme model. In this paper we will study the effect of replacing the baby Skyrme term with a vector meson term in a curved background, and see how the isolated solitons and popcorn transitions of finite-density configurations are affected. In addition, since full numerical field simulations are viable in two dimensions, we will use the low-dimensional vector meson model as a testing ground to check the validity of some common approximations used when studying the full Sakai-Sugimoto model.

\section{The holographic baby Skyrme and vector meson models}
We consider a $(D+2)$-dimensional spacetime with a warped metric
\begin{equation}\label{metric}
ds^2 = H(z)(-dt ^2+dx_1^2 + ... + dx_D^2) + \frac{1}{H(z)}dz^2
\end{equation}
with
\begin{equation}
H(z) = \left(1+\frac{z^2}{L^2}\right)^p \, .
\end{equation}
The warp factor, $H(z)$ depends only on the holographic coordinate $z$, and $L$ determines some curvature length scale which we will set to unity. We will choose the constant $p$ later.

The metric of the Sakai-Sugimoto model is recovered by choosing $D=3$ and $p=\frac{2}{3}$, which ensures that the spacetime has a conformal boundary as $z\rightarrow\infty$, and that the scalar curvature is negative and finite everywhere.

Following \cite{Bolognesi:2013jba}, we investigate a lower-dimensional toy model of the Sakai-Sugimoto model with $D=1$. To ensure a negative and finite scalar curvature we take $p$ to lie in the range $\frac{2}{5}\le p\le 1$, and for flat-space instanton approximations of the solitons to have finite energy we require the further restriction $p<\frac{2}{3}$. It is therefore convenient to make the choice $p=\frac{1}{2}$.

In \cite{Bolognesi:2013jba} solitons in the baby Skyrme model in the spacetime \eqref{metric} were studied. The action for this baby Skyrme (BS) model is
\begin{equation}\label{bsaction}
S_{BS} = \int \left( -\frac{1}{2}\partial_\mu\pmb{\phi}\cdot\partial^\mu\pmb{\phi} - \frac{\kappa^2}{4}(\partial_\mu\pmb{\phi}\times\partial_\nu\pmb{\phi})\cdot(\partial^\mu\pmb{\phi}\times\partial^\nu\pmb{\phi}) \right)\sqrt{H}\, dx\, dz\, dt
\end{equation}
where the first term is that of the $O(3)$-sigma model, and the second term is the baby Skyrme term with constant coefficient $\kappa^2$. Greek indices run over the bulk spactime coordinates $t$, $x$ and $z$.  The field $\pmb{\phi} = (\phi_1 , \phi_2 , \phi_3)$ is a three component unit vector. We will refer to $\pmb{\phi}$ as the pion field since the baby Skyrme model can be seen as a 2-dimensional analogue of the Skyrme model where the corresponding $\pmb{\phi}$ field plays the role of pions. The associated static energy of this model is
\begin{equation}\label{BSE}
E_{BS} = \frac{1}{2}\int\,\left( \frac{1}{\sqrt{H}}|\partial_x\pmb{\phi}|^2 + H^{3/2} |\partial_z\pmb{\phi}|^2 + \kappa^2\sqrt{H} |\partial_x\pmb{\phi}\times\partial_z\pmb{\phi}|^2 \right)\, dx\, dz\, .
\end{equation}

Here we will instead investigate solitons of an $O(3)$-sigma model coupled to a massive vector meson. The action for this vector meson (VM) model is given by
\begin{multline}\label{vmaction}
S_{VM} = \int \bigg( -\frac{1}{2}\partial_\mu\pmb{\phi}\cdot\partial^\mu\pmb{\phi} - \frac{1}{4}(\partial_\mu \omega_\nu - \partial_\nu \omega_\mu)(\partial^\mu \omega^\nu - \partial^\nu \omega^\mu)\\ - \frac{1}{2}M^2\omega_\mu\omega^\mu + g\omega_\mu B^\mu \bigg)\sqrt{H}\, dx\, dz\, dt
\end{multline}
where the pion field $\pmb{\phi} = (\phi_1 , \phi_2 , \phi_3)$ is again a three component unit vector and $\omega_\mu$ is a spacetime vector field with mass $M$. The fourth term, parametrised by the constant $g$, is a coupling between the $\omega_\mu$ field and the topological current
\begin{equation}\label{topcharge}
B^\mu = -\frac{1}{8\pi\sqrt{H}}\varepsilon^{\mu\alpha\beta}\pmb{\phi}\cdot (\partial_\alpha\pmb{\phi}\times\partial_\beta\pmb{\phi}) \, .
\end{equation}

We argue that this model is a low-dimensional analogue of the Sakai-Sugimoto model. The full model is written in terms of a $U(2)$ gauge field which can be decomposed into an abelian $U(1)$ and an $SU(2)$ field. Written in this way, the Chern-Simons term is a coupling between the $U(1)$ field and the topological current associated with the $SU(2)$ field. Noting that both pure $SU(2)$ theory in $(4+1)$ dimensions and the $O(3)$-sigma model in $(2+1)$ dimensions are both scale invariant, we can take $\pmb{\phi}$ and $\omega_\mu$ to be the low-dimensional analogues of the $SU(2)$ and $U(1)$ fields respectively. With this correspondence we see that the fourth term in \eqref{vmaction} is a low-dimensional analogue of the Chern-Simons term.

In this paper we will be concerned with static soliton solutions in this model. Looking at \eqref{topcharge} it is easy to see that for static solutions the spatial components of the topological current vanish ($B^i=0$), where the index $i$ runs over the spatial coordinates $x$ and $z$. These currents provide a source for the omega field, so we see that $\omega_i=0$ for static solutions. For notational convenience we will write $\omega\equiv\omega_0$ for the remainder of this paper. We will refer to $\omega$ as the vector meson field.

The associated static energy of this model is
\begin{equation}\label{totenergy}
E_{VM} = E_{\pmb{\phi}} + E_\omega
\end{equation}
with
\begin{align}
E_{\pmb{\phi}} &= \frac{1}{2}\int\,\left( \frac{1}{\sqrt{H}} |\partial_x\pmb{\phi} |^2 + H^{3/2} |\partial_z\pmb{\phi} |^2  \right)\, dx\, dz \label{Ephi}
\\
E_\omega &= \frac{1}{2}\int\,\left[ - \frac{1}{H^{3/2}}(\partial_x\omega)^2 - \sqrt{H}(\partial_z\omega)^2 - \frac{M^2\omega^2}{\sqrt{H}} + \frac{g\omega}{2\pi}\pmb{\phi}\cdot (\partial_x\pmb{\phi}\times\partial_z\pmb{\phi}) \right]\, dx\, dz \, . \label{Eomega}
\end{align}
For finite energy we then require $\pmb{\phi} \rightarrow (0,0,1)$ and $\omega\rightarrow 0$ as $x^2+z^2\rightarrow\infty$. This allows us to compactify space from $\mathbb{R}^2$ to $S^2$, meaning the pion field is now a map $\pmb{\phi}:S^2\rightarrow S^2$ with an associated winding number and topological charge
\begin{equation}
B = \int\, B^0 \, \sqrt{H}\, dx\, dz = -\frac{1}{4\pi}\int\,\pmb{\phi}\cdot(\partial_x\pmb{\phi}\times\partial_z\pmb{\phi})\, dx\, dz
\end{equation}
which defines the instanton number of the planar $O(3)$-sigma model. This topological charge we identify with the baryon number of the configuration.

By calculating the variation of the action \eqref{vmaction} with respect to the vector meson field we obtain the Euler-Lagrange equation
\begin{equation}\label{omegaEOM}
\frac{\delta E_{VM}}{\delta\omega} = \left( \frac{1}{H^{3/2}}\partial_{xx} + \sqrt{H}\partial_{zz} + \frac{1}{2}\frac{H^\prime}{\sqrt{H}}\partial_z - \frac{1}{\sqrt{H}}M^2 \right)\omega + \frac{g}{4\pi}\pmb{\phi}\cdot(\partial_x\pmb{\phi}\times\partial_z\pmb{\phi}) = 0 \, .
\end{equation}
Multiplying this equation by $\omega$ and integrating by parts in \eqref{Eomega} allows us to write
\begin{align}
E_\omega &= \frac{1}{2}\int\, \frac{g}{4\pi}\omega\pmb{\phi}\cdot(\partial_x\pmb{\phi}\times\partial_z\pmb{\phi})\, dx\, dz \label{simpleenergy}
\\
&= \frac{1}{2}\int\, \left( \frac{1}{H^{3/2}}(\partial_x\omega)^2 + \sqrt{H}(\partial_z\omega)^2 + \frac{M^2\omega^2}{\sqrt{H}} \right)\, dx\, dz \, .
\end{align}
We then see that $E_\omega\ge 0$ whenever the equation of motion for $\omega$ is satisfied and this, together with the inequality
\begin{equation}\label{Bog}
\left|\frac{1}{\sqrt{H}}\partial_x\pmb{\phi} \pm \sqrt{H}\pmb{\phi}\times\partial_z\pmb{\phi} \right|^2\ge 0\, ,
\end{equation}
yields the Bogomolny bound $E_{VM}\ge 4\pi |B|$.

By approximating the solution to \eqref{omegaEOM} using a leading order derivative expansion, we can see how the BS and VM models are related \cite{Foster:2009rw}. Neglecting derivatives of $\omega$ in \eqref{omegaEOM} yields the approximation
\begin{equation}\label{omegaapprox}
\omega\approx\frac{g\sqrt{H}}{4\pi M^2}\pmb{\phi}\cdot(\partial_x\pmb{\phi}\times\partial_z\pmb{\phi})
\end{equation}
which can be substituted into \eqref{simpleenergy} to obtain
\begin{align}
& E_\omega = \frac{1}{2}\int\, \frac{g^2\sqrt{H}}{16\pi^2 M^2}|\partial_x\pmb{\phi}\times\partial_z\pmb{\phi}|^2\, dx\, dz
\\
\Rightarrow\, & E_{VM} = \frac{1}{2}\int\,\left( \frac{1}{\sqrt{H}}|\partial_x\pmb{\phi}|^2 + H^{3/2} |\partial_z\pmb{\phi}|^2 + \frac{g^2\sqrt{H}}{16\pi^2 M^2} |\partial_x\pmb{\phi}\times\partial_z\pmb{\phi}|^2 \right)\, dx\, dz\, .
\end{align}
This is precisely the static energy of the BS model \eqref{BSE} upon identifying 
\begin{equation}\label{kappa}
\kappa = \frac{g}{4\pi M}\, .
\end{equation}
Note that the approximation used above becomes more accurate as $M$ and $g$ increase i.e. as the vector meson field becomes infinitely massive and its interaction becomes point-like. We recover the exact BS model in the limit
\begin{equation}
M, g\rightarrow \infty\, , \qquad \kappa=\frac{g}{4\pi M} = \mathrm{constant}\, .
\end{equation}
We will justify this approximation and discuss its validity in the following section.

\section{Solitons in the vector meson model}
In flat space ($H=1$) and in the absence of a vector meson ($\omega_\mu\equiv0$) the Bogomolny bound \eqref{Bog} is attained by the scale-invariant instanton solutions of the $O(3)$-sigma model. To write explicit solutions to this model it is convenient to consider Riemann sphere coordinates $W=\frac{\phi_1 + i\phi_2}{1-\phi_3}$, obtained by stereographic projection of the pion field $\pmb{\phi}$. Instanton solutions with finite $B>0$ are then given by rational functions $W(\zeta)$ of a complex coordinate $\zeta = x+iz$, where the required boundary conditions, $W\to\infty$ as $|\zeta|\to\infty$, imply the degree of the numerator of $W$ must be greater than the degree of the denominator. This leaves us with an instanton moduli space $\mathcal{M}_B$ of dimension $4B-1$, after considering the $U(1)$ symmetry associated with the phase of $W$.

The radially symmetric sigma-model instanton with topological charge $B$ centred at the origin is given by $W=(\zeta/\mu)^B$ where $\mu$ is a positive real constant which determines the arbitrary scale of the instanton. The associated pion field for this solution is given by
\begin{equation}\label{flatsigmaansatz}
\pmb{\phi} = (\sin{f}\cos{(B\theta)}, \sin{f}\sin{(B\theta)},\cos{f})
\end{equation}
where $(r,\theta)$ are polar coordinates in the $(x,z)$-plane and $f(r)$ is the radial profile function
\begin{equation}\label{profile}
f(r) = \cos{}^{-1}\left( \frac{r^{2B} - \mu^{2B}}{r^{2B} + \mu^{2B}} \right)\, .
\end{equation}

In the Sakai-Sugimoto model the 't Hooft coupling is taken to be large, which results in a small size for the Sakai-Sugimoto soliton compared to the length scale set by the curvature of spacetime. The analogous regime in the BS model is to take $\kappa$ small and positive, and it was shown that in this regime the soliton can be well approximated by a sigma model instanton. These instantons have an arbitrary size in flat space, but a preferred size can be obtained by considering the leading order contributions to the energy from the spacetime curvature and the baby Skyrme term interaction. Here we shall search for some parameter regime in $(g,M)$-space in the VM model analogous to this situation, and demonstrate that this regime is consistent with the approximations discussed above, and with the established relation between the VM and BS models.

Taking a flat-space ($H=1$) radial ansatz for the vector meson field, and a flat-space sigma instanton ansatz \eqref{flatsigmaansatz} for the pion field, we first find a parameter regime in which the approximation \eqref{omegaapprox} is valid. With these assumptions, the equation of motion \eqref{omegaEOM} for $\omega$ becomes
\begin{equation}
\frac{1}{r}\partial_r (r\partial_r\omega) = M^2\omega + \frac{g B^2\mu^{2B}}{\pi}\frac{r^{2B-2}}{(r^{2B}+\mu^{2B})^2}
\end{equation}
where $B$ is the topological charge of the solution and $\mu$ is the size of the sigma instanton. Assuming that $|\frac{1}{r}\partial_r (r\partial_r\omega)|\ll |M^2\omega|$ yields the approximation
\begin{equation}\label{omegaapprox2}
\omega\approx\frac{-g B^2\mu^{2B}}{M^2\pi}\frac{r^{2B-2}}{(r^{2B}+\mu^{2B})^2}\, .
\end{equation}
We then find
\begin{align}
\left|M^2\omega\right| &= \frac{g B^2\mu^{2B}}{\pi}\frac{r^{2B-2}}{(r^{2B}+\mu^{2B})^2}\, ,
\\
\left|\frac{1}{r}\partial_r (r\partial_r\omega)\right| &=
\frac{4gB^2\mu^{2B}}{M^2\pi}\frac{r^{2B-4}}{(r^{2B}+\mu^{2B})^4}\Gamma\, , 
\end{align}
where $\Gamma = \left| (B+1)^2 r^{4B} - 2(2B^2-1)\mu^{2B} r^{2B} + (B-1)^2\mu^{4B} \right|$. Therefore the inequality above is consistent if
\begin{equation}
\frac{4}{M^2(r^{2B}+\mu^{2B})^2}\left| (B+1)^2 r^{4B-2} - 2(2B^2-1)\mu^{2B} r^{2B-2} + (B-1)^2\frac{\mu^{4B}}{r^2} \right| \ll 1\, ,
\end{equation}
For $B=1$ the approximation \eqref{omegaapprox2} has a local minimum at the origin, and evaluating the above expression there yields the condition
\begin{equation}\label{Mbound}
M \gg \frac{2\sqrt{2}}{\mu}\, .
\end{equation}
For other topological charges we can find a similar condition by evaluating the above expression at the extrema of the radial ansatz. We have therefore found that if we seek a regime in which the soliton size is small compared to the curvature scale, and in which derivatives of $\omega$ are small, we require a large mass for the vector meson field.

We will now insert this approximation into the curved-space static energy and look at the terms coming from the first-order expansion of the warp factor, $H(z)$. We shall also re-insert the explicit dependence of $H$ on $p$ so that we can clearly see which terms arise from the curvature of spacetime and which terms arise from interactions between the pion and vector meson fields.

Expanding in the curvature, we have
\begin{equation}\label{approxx}
E_{\pmb{\phi}} = \frac{1}{2}\int\,\left( |\partial_x\pmb{\phi}|^2 + |\partial_z\pmb{\phi}|^2\right)\, dx\, dz + \frac{p}{2}\int\, \left( -|\partial_x\pmb{\phi}|^2 + 3|\partial_z\pmb{\phi}|^2 \right) z^2\, dx\, dz + \cdots
\end{equation}
The first term in this expansion is the usual flat-space sigma instanton energy, which has the well-known value of $4\pi B$. The second term is some contribution arising from the curvature of spacetime and, after changing to polar coordinates and making the change of variable $y=(r/\mu)^2$, its leading order contribution can be calculated as
\begin{equation}\label{integral1}
E_{\pmb{\phi}} = 4\pi B + p\pi B^2\mu^2\int_0^\infty\,\frac{y^B}{(y^B+1)^2}\, dy+\cdots
\end{equation}
We see that the influence of the curvature term, therefore, is to reduce the size of the instanton.

Using a similar expansion, the leading order contribution from the vector meson part of the energy can be calculated as
\begin{equation}\label{integral2}
E_{\omega} = \frac{g^2 B^4}{2\pi M^2\mu^2}\int_0^\infty\,\frac{y^{2(B-1)}}{(y^B+1)^4}\, dy+\cdots
\end{equation}
In contrast to the curvature term, we see that the interaction with the vector meson field resists the shrinking of the instanton.

For $B=2$ we can evaluate these integrals explicitly and obtain the leading order correction to the energy as
\begin{equation}
E_{VM} = 8\pi + p\pi^2\mu^2 + \frac{g^2}{4M^2\mu^2} + \cdots
\end{equation}
which is minimised when
\begin{equation}\label{solsize}
\mu = \left( \frac{g}{2\pi M\sqrt{p}} \right)^{1/2}\, .
\end{equation}
We see that, as claimed, the radial flat-space approximations are valid for large $g$ and large $M$, and in this regime the instanton solutions in the VM model closely resemble those in the BS model. In fact, substituting the expression for $\kappa$ in \eqref{kappa} we find the preferred instanton size is $\mu = \sqrt{2\kappa}/p^{1/4}$, which is in exact agreement with the analysis performed in \cite{Bolognesi:2013jba} for the BS model.

\begin{figure}[t]
	\centering
	\includegraphics[angle=270, width=0.55\textwidth]{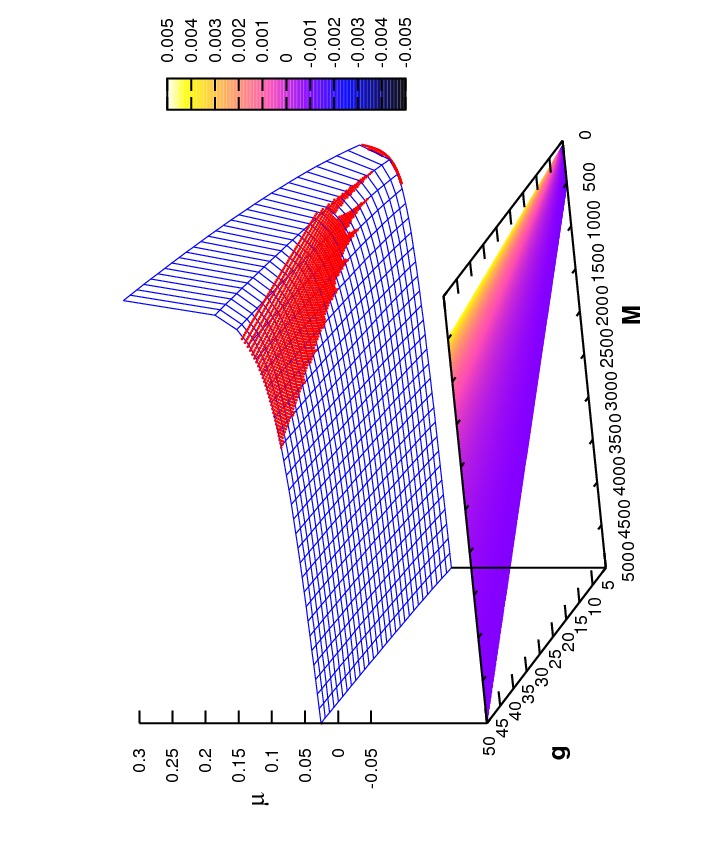}
	\caption{A plot of the size $\mu$ against the model parameters $g$ and $M\in[10g, 100g]$ for the $B=1$ soliton within the instanton approximation. The red lines give the numerical data, and the blue surface is a numerical fit to the data of the form $\mu = c\sqrt{g/(4\pi M)}$ with constant $c=0.92$. The contour plot at the base gives the difference between these two surfaces, and is seen to be small.}
	\label{fig1}
\end{figure}

A similar calculation can be performed for all $B\ge 2$ to show that the preferred instanton size is proportional to $\sqrt{g/M}$ in the regime of large $g$ and large $M$. However, the integral expressions in \eqref{integral1} and \eqref{integral2} do not converge for $B=1$. A similar problem was encountered in \cite{Bolognesi:2013jba} when studying the BS model, which arises since the quadratic approximation for $H(z)$ does not capture the slow decay behaviour of the $B=1$ solution accurately enough. In analogy with this previous paper, we can insert the instanton ansatz of size $\mu$ into the expression for the energy \eqref{totenergy} and minimise this numerically to find $\mu$ as a function of $g$ and $M$. The result for $p=\frac{1}{2}$ is displayed in Figure~\ref{fig1} for $g\in [5,50]$ and $M\in[10 g, 100 g]$ (roughly corresponding to $\kappa\in [0.01, 0.001]$ after the identification \eqref{kappa}), where the red lines represent the numerical results and the blue surface is a fit to the data of the form $c\sqrt{g/(4\pi M)}$ with constant $c=0.92$. The computation suggests that, in this regime, the size of the $B=1$ soliton still has an approximate size $\mathcal{O}(\sqrt{g/M})$, in agreement with the analytic calculation for the $B=2$ above.

In this two-dimensional model it is possible to obtain full numerical static solutions to the non-linear field equations. All of the computations performed here are with $p=\frac{1}{2}$ and with the ratio $\frac{g}{M}=0.1$ kept constant, so that in the large $g$, large $M$ limit we expect solutions to look like those of the BS model with parameter $\kappa=\frac{g}{4\pi M}\approx0.01$. In this limit the size of the soliton should be small compared to the curvature scale set by the metric. Numerical computations have been performed by minimising the energy \eqref{totenergy} using second-order accurate finite difference approximations for spatial derivatives on a lattice with spacings $\Delta x=0.005$, $\Delta z=0.005$ and $2000\times 2000$ numerical grid points. The energy minimisation algorithm is a modified gradient flow mechanism.

\begin{figure}[p]
	\centering
		\includegraphics[width=0.40\textwidth]{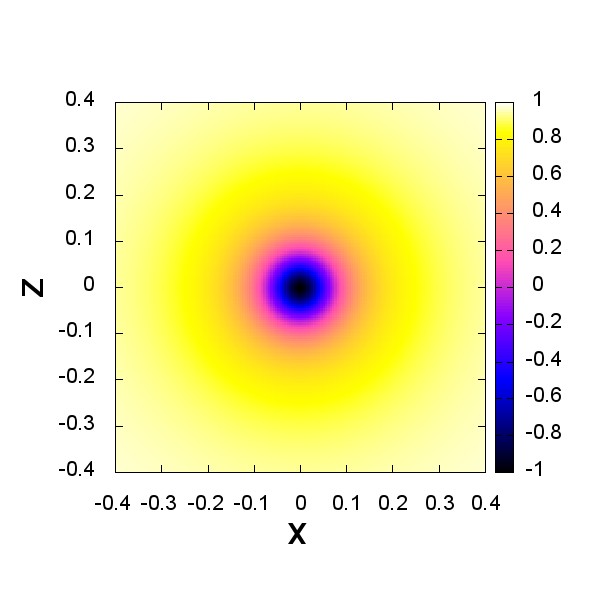}
		\includegraphics[width=0.40\textwidth]{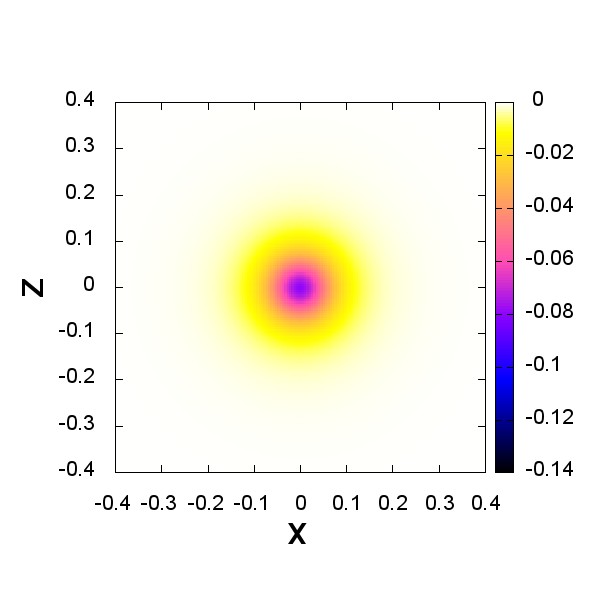}
	\caption{Plots of the numerically calculated 1-soliton fields for the $M=50$ VM model. $\phi_3$ is shown on the left and $\omega$ is shown on the right.}
	\label{chargeones}
\end{figure}

The results of our field theory computations for the $B=1$ soliton are shown in Figure~\ref{chargeones}. The left image shows the $\phi_3$ component of the pion field, as a pictorial representation of the soliton, for parameter values $M=50$ and $g=5$. The right image shows the corresponding vector meson field $\omega$. The left image is almost indistinguishable from the numerical calculation of the $B=1$ soliton in the BS model with parameter value $\kappa=0.1/(4\pi)\approx0.008$. This is confirmed by looking at the energies of each solution: the energy of the $B=1$ VM soliton with $M=50$ is given as $E=4\pi\times 1.0107$, while the corresponding BS soliton has energy $E=4\pi\times 1.0115$. Numerical computations for smaller $M$ and $g$ show that, as the parameter values decrease, the size of the VM soliton appears to decrease.

\begin{figure}[p]
	\centering
	\begin{subfigure}[b]{\textwidth}
		\centering
		\includegraphics[width=0.40\textwidth]{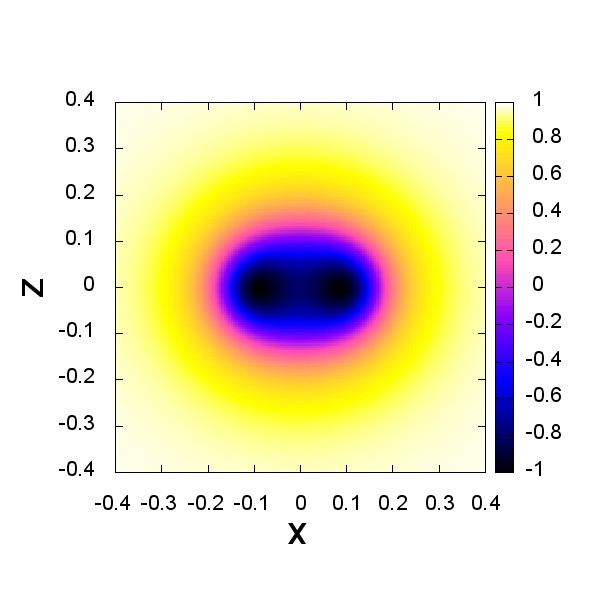} 
		\includegraphics[width=0.40\textwidth]{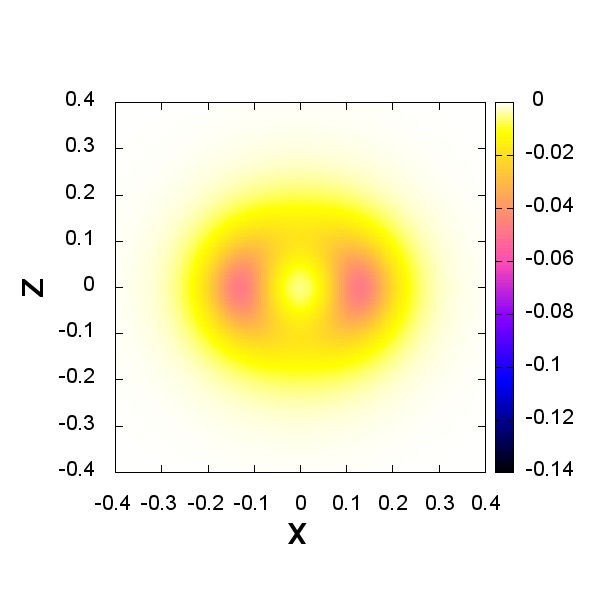}
		\label{VM250}
	\end{subfigure}
	\hfill\\
	\begin{subfigure}[b]{\textwidth}
		\centering
		\includegraphics[width=0.40\textwidth]{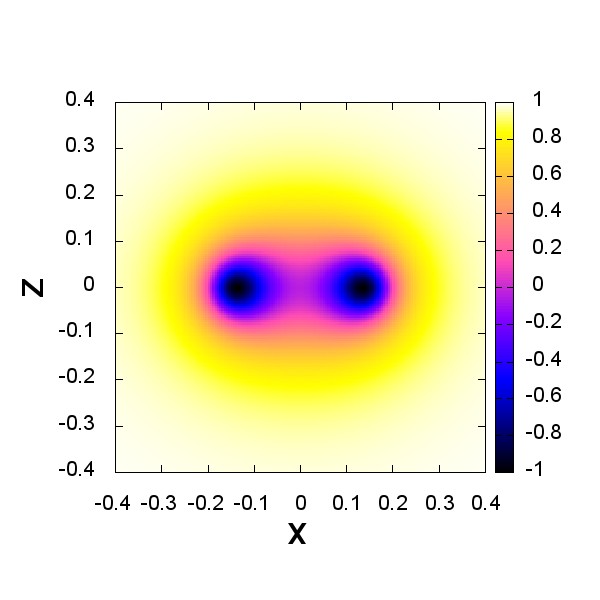}
		\includegraphics[width=0.40\textwidth]{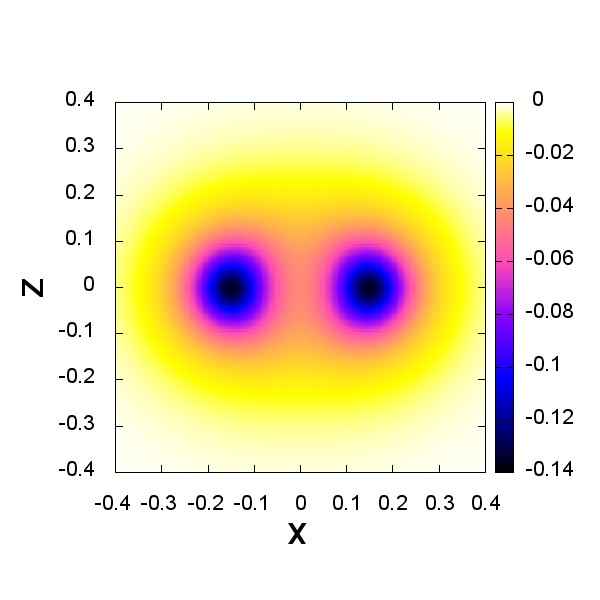}
		\label{VM215}
	\end{subfigure}
	\caption{Plots of the numerically calculated 2-soliton fields for the $M=50$ (top) and $M=15$ (bottom) VM model. $\phi_3$ is shown in the left column and $\omega$ is shown in the right column.}
	\label{chargetwos}
\end{figure}

The numerical results for charge $B=2$ are shown in Figure~\ref{chargetwos}. For large parameter values we again find that the solitons of the VM and BS model are almost indistinguishable: an energy comparison yields $E=8\pi\times 1.0072$ for the VM model soliton with $M=50$, and $E=8\pi\times 1.0075$ for the corresponding BS soliton. In these cases the $B=2$ soliton resembles two $B=1$ solitons separated by a small amount in the non-holographic direction. However, as we approach the region in which the approximation \eqref{approxx} breaks down we find qualitatively different behaviour, in which the $B=1$ components of the $B=2$ soliton begin to separate. For $M<20$ the apparent separation between these components increases dramatically, as shown in the final image in Figure~\ref{chargetwos}. We also see that as the parameters in the model decrease, the $\omega$ fields become more sharply localised around their centres. This is consistent with the analysis above where we found small derivatives in $\omega$ implied a large value of $M$.

Solitons with higher topological charge have also been numerically computed, and are found to resemble chains of solitons separated in the non-holographic direction. This is to be expected due to the presence of the curvature in the holographic direction.

In \cite{Kaplunovsky:2012gb} a zigzag arrangement of holographic baryons was predicted in finite density configurations in the Sakai-Sugimoto model, and in \cite{Bolognesi:2013jba} it was argued that, for the low-dimensional BS analogue of the Sakai-Sugimoto model, a zigzag arrangement could only be found if the preferred separation between pairs of solitons was much greater than the size of a single soliton. We have seen that, at low parameter values, the $B=1$ VM solitons decrease in size and the $B=2$ solitons resemble more widely separated $B=1$ solitons, so it would be interesting to explore finite density chains of solitons in the VM model for small values of $g$ and $M$. However, at these parameter values numerical solutions become more difficult to obtain (due to large derivatives of the $\omega$ field), so we will instead study a regime in which the parameters are ``small, but not too small''.

\section{Finite density chains: New phases of popcorn}
As explained in\cite{Bolognesi:2013jba}, analytical and numerical solutions of solitons at finite density in the Sakai-Sugimoto model are currently unavailable. Despite various levels of approximation being applied, even the relevant flat-space instanton approximations are not known explicitly after imposing periodic boundary conditions in multiple directions. However, the lower-dimensional analogue we have been studying so far gives us the advantage of being able to numerically compute soliton solutions, and we shall compare our field theory simulations to the numerical and analytical results in \cite{Bolognesi:2013jba}. In particular, we will find the appropriate popcorn phase transition in the VM model, and compare it both to the BS model, and to predictions made in \cite{Kaplunovsky:2012gb} and \cite{Kaplunovsky:2015zsa} about finite-density baryons in the full Sakai-Sugimoto model.

In order to numerically compute solitons at finite density we restrict our numerical grid in the non-holographic direction to the range $-L\le x\le L$ and impose suitable boundary conditions for the fields at $x=\pm L$. In the following simulations we will always place periodic boundary conditions on $\phi_3$ and $\omega$, but may place periodic or anti-periodic boundary conditions on $\phi_1$ and $\phi_2$. There is still an integer topological charge, $B$, defined by integrating the $\mu=0$ component of \eqref{topcharge} in the range $(x,z)\in [-L,L]\times (-\infty,\infty)$, and this allows us to define a finite baryon density $\rho = B/2L$.

Initial conditions for numerical simulations are constructed as follows. First, by writing the flat-space instanton of charge $B=1$ using Riemann sphere coordinates, we can superpose a number of instanton solutions together using the product ansatz:
\begin{equation}\label{productansatz}
(W_1(\zeta), W_2(\zeta))\to W(\zeta)=\frac{W_1 W_2}{W_1 + W_2}\, .
\end{equation}
This has the property that $W$ vanishes wherever $W_1$ or $W_2$ vanishes, and that far away from $W_2$ the solution looks like $W\approx W_1$ (and vice-versa). Therefore this product ansatz gives us a field configuration with the correct topological properties as a superposition of solitons at the positions of $W_1$ and $W_2$. Superposing a large number of such solitons with positions in the range $x\in[-nL,nL]$ for some $n\in\mathbb{Z}$ using the flat-space instanton ansatz from \eqref{flatsigmaansatz} and \eqref{profile} can then give us an approximation to a periodic chain of solitons in the range $x\in[-L,L]$. An initial condition for the $\omega$ field can then be generated using \eqref{omegaapprox}. It should be noted that with this construction each soliton in the chain can be given an independent phase shift $\chi$ corresponding to a shift $\theta\to\theta +\chi$ in the ansatz \eqref{flatsigmaansatz}.

Inspired by the results found in \cite{Bolognesi:2013jba}, we will begin our investigation by searching for two specific forms of finite-density chains by imposing certain symmetries, before going on to find the global energy minima for periodic solutions in the VM model.

\begin{figure}[p]
	\centering
	\includegraphics[width=0.7\textwidth]{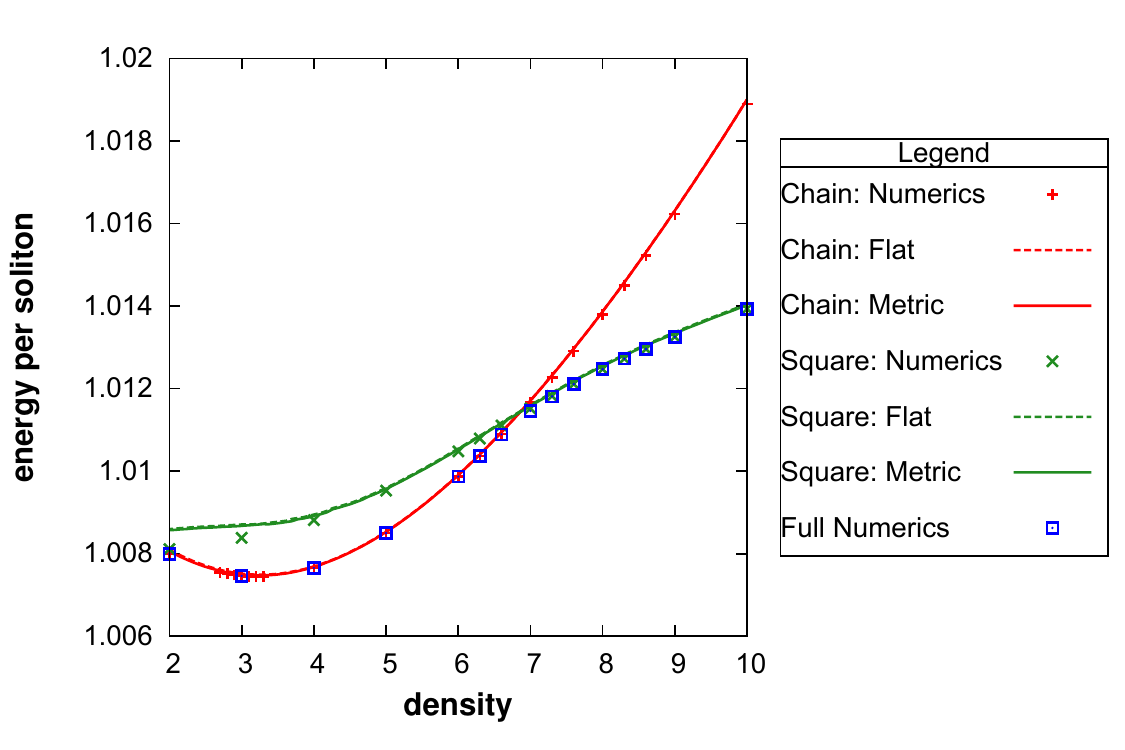}
	\caption{A plot of $E/(4\pi B)$ against density $\rho$ for the forced chain and square configurations, as well as the energy minimising solutions, for $M=50$. The data points represent the results of numerical simulations, whereas the curves represent the results of applying various semi-analytical approximations. The dotted lines are the results from approximating spacetime as flat, and the solid curves are the results from performing a first-order expansion of the metric factor $H(z)$.}
	\label{chainenergies}
\end{figure}

The first chain to consider is a straight, single chain of solitons. Flat-space instantons of equal size separated in the non-holographic direction can be shown to be in the most attractive channel when they are exactly out of phase (i.e.\ when the solitons have a relative phase shift of $\pi$), so we numerically seek these solutions by placing a single $B=1$ ansatz on our numerical grid and imposing anti-periodic boundary conditions on $\phi_1$ and $\phi_2$. The datapoints in Figure~\ref{chainenergies} marked with a $+$ represent the energy per soliton for numerical solutions of this type.

Secondly, we consider a two-layer square chain by placing a pair of solitons (with a relative phase shift of $\pi$) separated in the holographic direction, with anti-periodic boundary conditions imposed on $\phi_1$ and $\phi_2$ at $x=\pm L$. Symmetry constraints can force the resultant energy minima to form a regular square configuration of solitons (as observed in the BS model in \cite{Bolognesi:2013jba}), although it can also allow configurations that resemble zig-zags or single chains of $2$-solitons. The data points in Figure~\ref{chainenergies} marked with a $\times$ represent the energy per soliton for numerical solutions of this type.

\begin{figure}[p]
	\centering
	\begin{subfigure}[b]{\textwidth}
		\centering
		\includegraphics[width=0.24\textwidth]{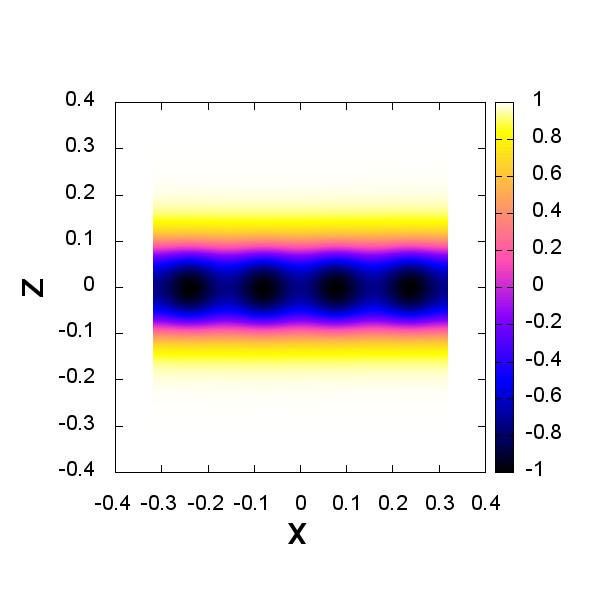}
		\includegraphics[width=0.24\textwidth]{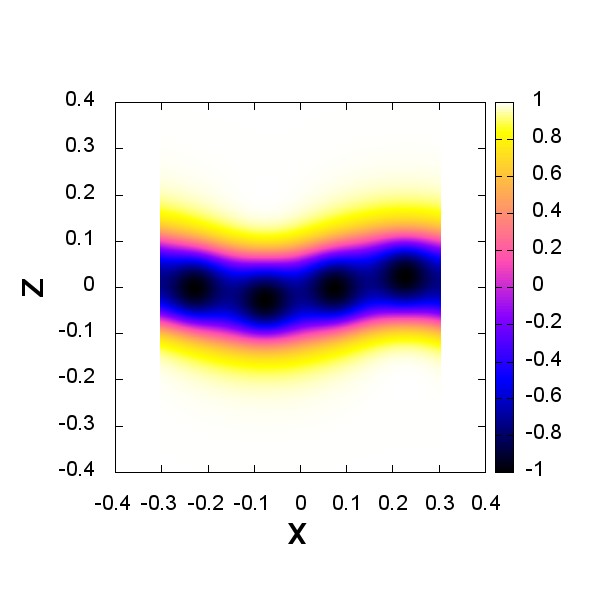}
		\includegraphics[width=0.24\textwidth]{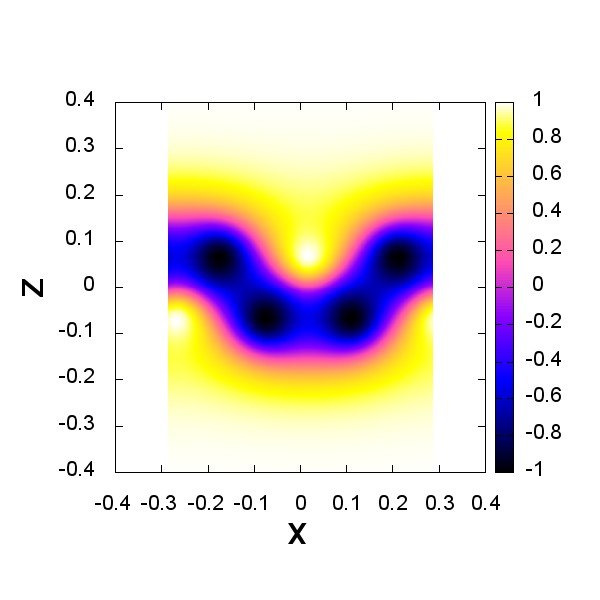}
		\includegraphics[width=0.24\textwidth]{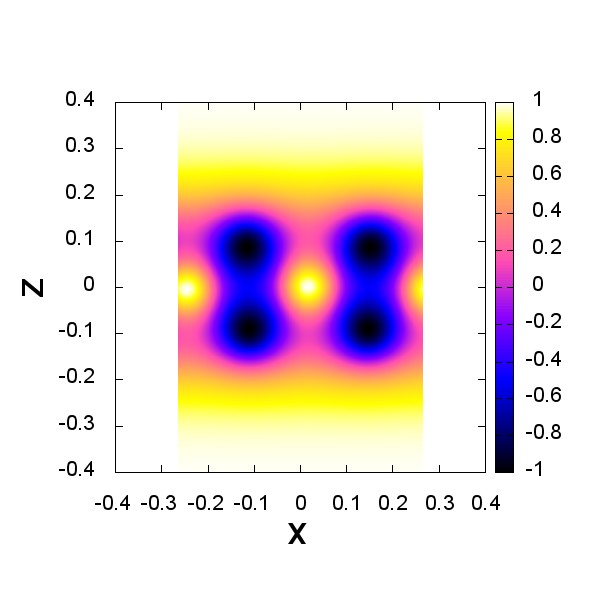}
		\label{wiggle1}
	\end{subfigure}
	\hfill\\
	\begin{subfigure}[b]{\textwidth}
		\centering
		\includegraphics[width=0.24\textwidth]{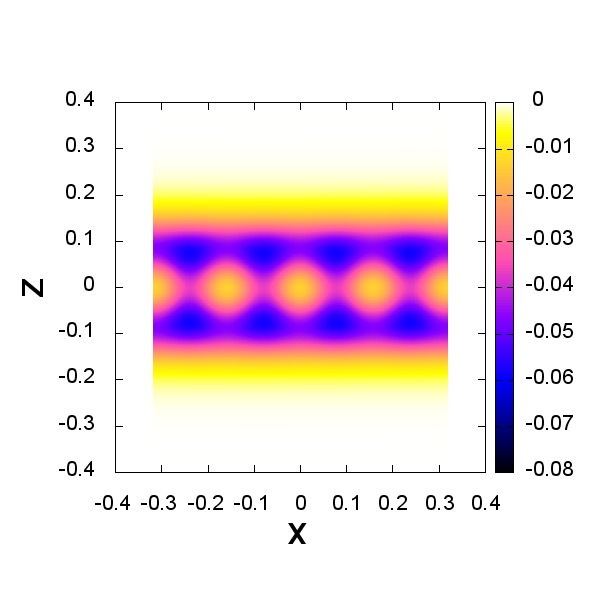}
		\includegraphics[width=0.24\textwidth]{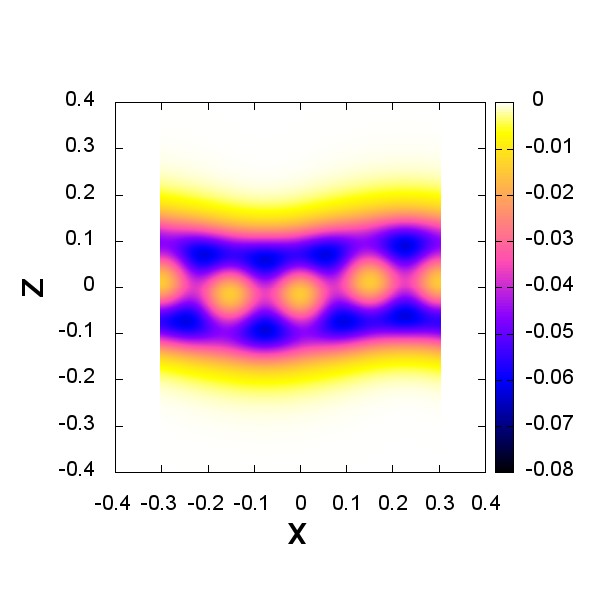}
		\includegraphics[width=0.24\textwidth]{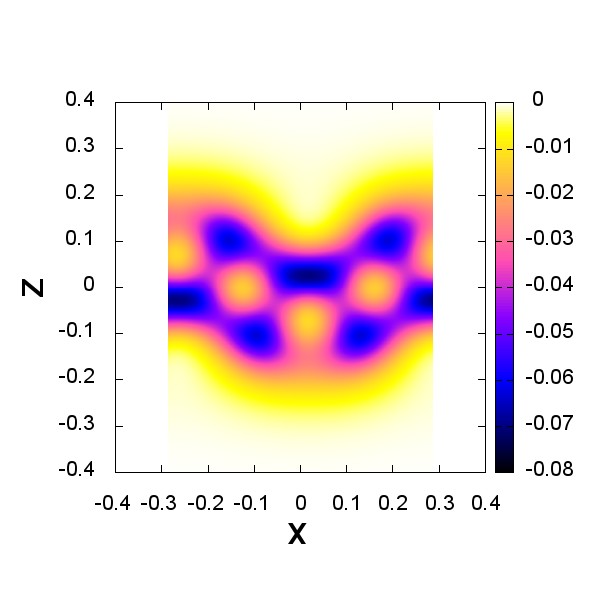}
		\includegraphics[width=0.24\textwidth]{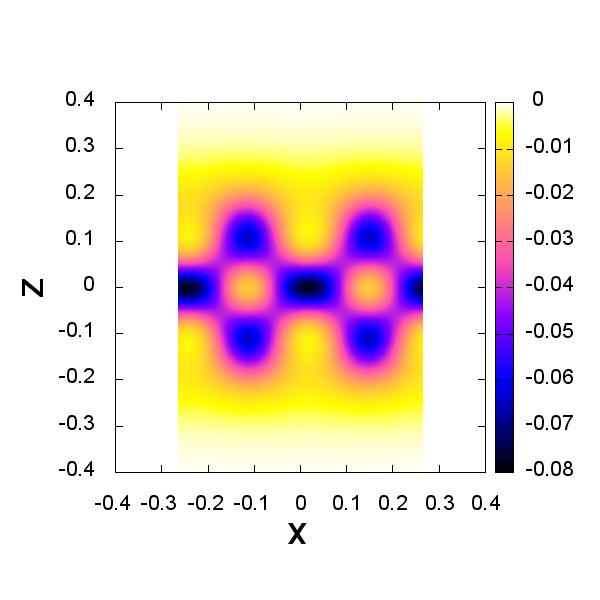}
		\label{wiggle2}
	\end{subfigure}
	\caption{The top row shows plots of $\phi_3$ for the energy minima of the VM model finite density chains with $M=50$ for densities $\rho=6.3$ (chain), $6.6$ (wave), $7$ (wiggle) and $7.6$ (square), with density increasing from left to right. The bottom row shows the associated $\omega$ fields.}
	\label{chains}
\end{figure}

Finally we consider initial conditions with four solitons placed in our grid in a line in the non-holographic direction, with periodic boundary conditions on $\phi_1$ and $\phi_2$. By itself there would be a reflection symmetry $z\to -z$ which could force solutions to resemble those of the straight chains above, so we perturb the fourth soliton in the chain by a small amount in the  holographic direction to break the symmetry. The symmetries of the grid here allow the potential for the initial conditions to relax to configurations resembling either the straight chains or the square chains, while also providing an opportunity to relax to other configurations with less symmetry. Figure~\ref{chains} shows $\phi_3$ and $\omega$ for different densities with $M=50$, and the data points in Figure~\ref{chainenergies} marked with a $\square$ represent the energy per soliton for the numerical solutions. We see that at low and high densities respectively we can recreate the straight and square configurations described above, lending confidence that those are indeed the energy minima at these densities.

More interestingly we find that there exist configurations of solitons in the VM model that were not found in \cite{Bolognesi:2013jba}. For densities $\rho\in [6.3,7.6]$ (i.e. around the transition from the chain to the square solutions) we first observe a second-order transition from a chain to a period-4 wave. This is followed by a first-order transition to a wiggle, which then undergoes a second-order transition to the two-layered square configuration. This is in contrast with the results anticipated in \cite{Kaplunovsky:2012gb} and \cite{Kaplunovsky:2015zsa}, the former predicting a first-order popcorn transition from a chain to a two-layered zig-zag, and the latter suggesting a second-order transition from a chain to a period-3 wave followed by a first-order transition to the square configuration. Imposing period-3 boundary conditions in this low-dimensional model yielded solutions that were not energetically favourable to the period-4 waves, suggesting that the anticipated period-3 waves are not global energy minima. While the numerical investigations performed here are not definitive, the evidence suggests that popcorn transitions in these models may be more subtle than previously anticipated.

The results in both studies are, nonetheless, similar, and the differences could be due to the lower dimensionality of the VM model studied here: in \cite{Kaplunovsky:2015zsa} chain solutions were generated in the 4-dimensional space by imposing artificial curvature in the other spatial directions, whereas in the 2-dimensional VM model studied here there are no extra spatial dimensions.

Numerical field calculations for finite-density solitons were also performed for $M=20$ and $g=2$. These parameter values are very close to limits of the bounds implied by \eqref{Mbound} and \eqref{solsize}. Qualitatively similar results were obtained to the solutions with parameter values $M=50$ and $g=5$, although investigation of smaller parameter values beyond those bounds proved numerically difficult due to the presence of large derivatives in the $\omega$ field.

The numerical differences in energy between the wave and wiggle configurations compared to the chain and square configurations are incredibly close, differing by roughly $O(10^{-5})$. In light of this, it was instructive to return to the BS model and see if these configurations did, in fact, occur but were missed due to this similarity. After investigation it was found that the BS model does, indeed, exhibit these configurations around the transition from the chain to the square.

\section{Semi-analytical methods of baryonic popcorn}
Another advantage of the VM model over the BS model is that the $\omega$ field provides a low-dimensional analogue of the $U(1)$ gauge field in the full Sakai-Sugimoto model. This means we can use the VM model to test the validity of certain assumptions that are often made to simplify calculations in the full model. In \cite{Kaplunovsky:2015zsa} a number of assumptions regarding the fields in the Sakai-Sugimoto model are made in order to make analytical progress. These assumptions all rely on the solitons of the model being much smaller than the length scale set by the spacetime curvature, in which case it is argued that we can approximate spacetime as flat, and the fields by their flat-space counterparts. The $SU(2)$ gauge field is approximated by a flat-space self-dual Yang-Mills instanton, allowing techniques using ADHM constructions to be used to generate solutions. The $U(1)$ gauge field is approximated by constraining it to satisfy its flat-space equations of motion. In addition, the metric factors due to the curvature of spacetime are expanded as a leading order Taylor series.

Since we have been able to obtain full numerical solutions in our low-dimensional model, we can apply analogues of these approximations to test their validity in the VM model. Focussing on the forced chain and square configurations allows us to obtain analytical approximations for the pion fields, and perform numerical minimisations to obtain the corresponding $\omega$ fields.

For the chain configurations we can write a flat-space instanton solution (in Riemann sphere coordinates) as
\begin{equation}
W(\zeta)=\frac{\sin{(\pi\rho\zeta)}}{\mu\pi\rho}\, .
\end{equation}
This has zeroes whenever $x = \frac{n}{\rho}$, $n\in\mathbb{Z}$ and obeys the symmetry relation $W(\zeta + \rho^{-1}) = -W(\zeta)$, so describes a chain of instantons at density $\rho$ where neighbouring instantons are exactly out of phase with each other. To obtain the square configurations we can simply take a product ansatz of the form \eqref{productansatz} with the choice 
\begin{equation}
W_1 = \frac{\sin{(\pi\rho(\zeta-i\delta)/2)}}{\mu\pi\rho}\,,\qquad W_2 = \frac{-\sin{(\pi\rho(\zeta+i\delta)/2)}}{\mu\pi\rho}
\end{equation}
to obtain a a pair of chains separated by a distance $\delta$ in the holographic direction.

Requiring that the $\omega$ field satisfies its flat-space equation of motion is equivalent to setting $H(z)=1$ in \eqref{omegaEOM}:
\begin{equation}\label{omegaEOMflat}
(\nabla^2 - M^2)\omega=-\frac{g}{4\pi}\pmb{\phi}\cdot(\partial_x\pmb{\phi}\times\partial_z\pmb{\phi})\, .
\end{equation}
Combining this equation with the instanton approximation above allows us to perform numerical minimisation of the energy \eqref{totenergy} over the parameters $\mu$ and/or $\delta$ for a large number of densities, where at each value of $\{\rho, \mu, \delta\}$ we solve \eqref{omegaEOMflat} numerically using a successive over-relaxation method with red-black ordering. The minimisation in parameter space was performed with a golden section search method (in the 1-parameter case) and a Nelder-Mead algorithm (in the 2-parameter case). The results of these investigations are shown by the curves in Figure~\ref{chainenergies}. As can clearly be seen, the energies associated with these approximations are very close to the full numerical values, lending confidence to the credibility of these assumptions. In addition, the effect of Taylor-expanding the metric factor $H(z)$ to leading order in the energy density was also investigated. The curves produced are also shown in Figure~\ref{chainenergies}, although it is difficult to see them since they essentially overlap with the flat-space approximation. This suggests that Taylor-expanding the metric is also a reasonable assumption in this regime of the model.

\section{Conclusions}
In this paper we have investigated an alternative low-dimensional analogue of the Sakai-Sugimoto model to describe holographic baryons, namely an $O(3)$-sigma model stabilised by vector mesons in a warped spacetime. We have found solitons, multi-solitons and finite density solitons in this model and compared them to solitons in the baby Skyrme model and to instanton approximations, and found all three to be similar given certain parameter constraints. We have also shown that solitons in the vector meson model can differ qualitatively when these parameters are allowed to violate such constraints.

In addition, we have used the vector meson model to investigate the low-dimensional analogue of the baryonic popcorn phenomenon, and found a set of new phase transitions (although such phase transitions are present in the baby Skyrme model, but were previously overlooked). We have used the vector meson model to test common approximations for the low-dimensional analogue of the $U(1)$ gauge field of the Sakai-Sugimoto model, and found that approximating the pion and vector meson fields by flat-space solutions and using a leading-order expansion of the metric are both good approximations.

When analysing the instanton approximations for finite-density solitons, only a very restricted set of instanton configurations were considered. It might be of interest to explore this area more fully to investigate the baryonic popcorn phase transition, by numerically minimising the energy of solutions where more instanton moduli (i.e. their positions and phases) are also allowed to vary. It may also be of interest to apply the various approximations tested here to high-density lattices of solitons in the full Sakai-Sugimoto model. Finally, although it is computationally much more difficult, it would be interesting to see how the phenomena observed in the vector meson model change for very small parameter values, since in this regime the model is no longer well-approximated by the baby Skyrme model.

\section{Acknowledgements}
This work is supported by an STFC PhD studentship. I would also like to thank my supervisor, Paul Sutcliffe for his support and useful discussions.

\bibliography{../../Papers/master.bib}
\bibliographystyle{../../Papers/JHEP}

\end{document}